%% file: panter_spectrograph_march19draft.tex
\documentclass[12pt]{spieman}  
\usepackage{amsmath}  
\usepackage[]{graphicx}
\usepackage{setspace}
\usepackage{tocloft}
\usepackage{color}
\usepackage{soul}
\usepackage{subcaption}
\usepackage{subcaption}
\usepackage[compact]{titlesec}
\usepackage{upgreek}
\usepackage{cleveref}
\crefname{section}{§}{§§}
\Crefname{section}{§}{§§}


\title{Performance Testing of a Novel Off-plane Reflection Grating and Silicon Pore Optic Spectrograph at PANTER} 

\author{Hannah Marlowe,\supscr{a}  Randall L. McEntaffer,\supscr{a} Ryan Allured,\supscr{b} Casey DeRoo,\supscr{a} Drew M. Miles,\supscr{a} Benjamin D. Donovan,\supscr{a} James H. Tutt,\supscr{a} Vadim Burwitz,\supscr{c} Benedikt Menz,\supscr{c} Gisela D. Hartner,\supscr{c} Randall K. Smith,\supscr{b} Ramses G\"{u}nther,\supscr{d} Alex Yanson,\supscr{d} Giuseppe Vacanti,\supscr{d} Marcelo Ackermann\supscr{e}}

\affiliation{\supscrsm{a}The University of Iowa Department of Physics $\&$ Astronomy, 210 Van Allen Hall, Iowa City, IA, USA\\
\supscrsm{b}Harvard-Smithsonian Center for Astrophysics, 60 Garden Street, Cambridge, MA, USA\\
\supscrsm{c}MPI f\"{u}r extraterrestrische Physik, Giessenbachstrasse 1, D-85748 Garching, Germany\\
\supscrsm{d}cosine Science \& Computing BV, J.H. Oortweg 19, 2333 CH Leiden, The Netherlands\\
\supscrsm{e}cosine Research BV, J.H. Oortweg 19, 2333 CH Leiden, The Netherlands\\}

\cftpagenumbersoff{figure}
\cftpagenumbersoff{table} 
\begin{document} 

\maketitle 

\begin{abstract}
An X-ray spectrograph consisting of radially ruled off-plane reflection gratings and silicon pore optics was tested at the Max Planck Institute for extraterrestrial Physics PANTER X-ray test facility. The silicon pore optic (SPO) stack used is a test module for the Arcus small explorer mission, which will also feature aligned off-plane reflection gratings. This test is the first time two off-plane gratings were actively aligned to each other and with a SPO to produce an overlapped spectrum. The gratings were aligned using an active alignment module which allows for the independent manipulation of subsequent gratings to a reference grating in three degrees of freedom using picomotor actuators which are controllable external to the test chamber. We report the line spread functions of the spectrograph and the actively aligned gratings, and plans for future development.
\end{abstract}

\keywords{Diffraction, gratings, grazing incidence, holography, x-rays}

{\noindent \footnotesize{\bf Address all correspondence to}: Hannah Marlowe, University of Iowa, Physics and Astronomy, 210 Van Allen Hall, Iowa City, Iowa, USA, 52242; Tel: +1 319-355-1835; Fax: +1 319-335-1753; E-mail:\linkable{hannah-marlowe@uiowa.edu} }


\section{Introduction}
\label{sect:intro}  

Arcus\cite{Smith14} is a proposed X-ray spectrograph to be installed on the International Space Station. This spectrograph consists of silicon pore optics (SPO\cite{Beijersbergen04}) and blazed, radially ruled off-plane reflection gratings optimized for the feature-rich soft X-ray regime. The mission utilizes the SPO being developed for ESA's Athena mission by cosine Research and reflection gratings being developed in collaboration between the University of Iowa and MIT/Lincoln Labs, and the mission . The Arcus mission will answer key science questions related to structure formation in the Universe, supermassive black hole feedback, and stellar life cycles. To meet its science objectives, Arcus will have a resoltuion of $\lambda/\Delta \lambda$~\textgreater~2500 and effective area \textgreater~600~cm$^{2}$ in the critical science bandpass around the O VII and O VIII lines (22.6 -- 25~\AA~). The mission will have a minimum resolution and effective area of $\lambda/\Delta \lambda$~\textgreater~1300 and \textgreater~130~cm$^{2}$ over the entire bandpass (8 -- 52~\AA~) with $\lambda/\Delta \lambda$ reaching $\sim$3000 at the longest wavelengths.

Performance testing of aligned off-plane reflection gratings with a SPO module was carried out at the PANTER\cite{Burwitz13} test facility of the Max Planck Institute for extraterrestrial Physics (MPE) in October 2014. During the tests, a radially ruled, off-plane reflection grating was aligned to the SPO test module. A second grating was then actively aligned to the first reference grating. This test was the first time that off-plane diffraction gratings were aligned with an SPO, and that two off-plane gratings were aligned to one another \textit{in situ}. This paper describes the experimental setup and results of the test campaign, specifically the comparison between the line spread functions of the SPO module and the first order Mg-K line from the aligned gratings. The components of the spectrograph are detailed in \S \ref{sect:telescope}, an overview of the alignment procedure is given in \S \ref{sect:exp}, CCD image reduction steps and measured line widths are presented in \S \ref{sect:reduction}--\ref{sect:results}, and discussion of the results in \S \ref{sect:disc}.

\section{Spectrograph Assembly}
\label{sect:telescope}
\begin{figure}[t]
   \begin{center}
   \includegraphics[width=.8\textwidth]{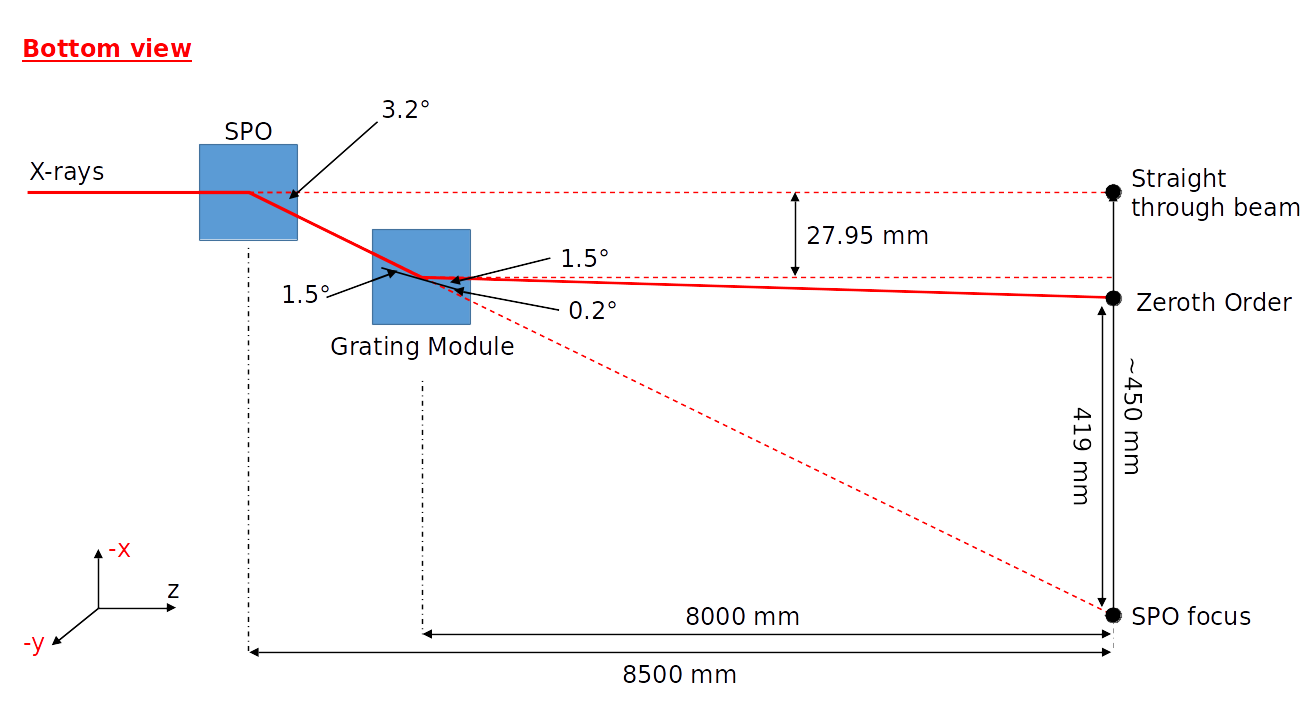} 
   \end{center}
   \caption 
   { \label{fig:sch} Bottom view of the test chamber SPO and grating integration.} 
   \end{figure}

\begin{figure}[t]
   \begin{center}
   \includegraphics[width=.7\textwidth]{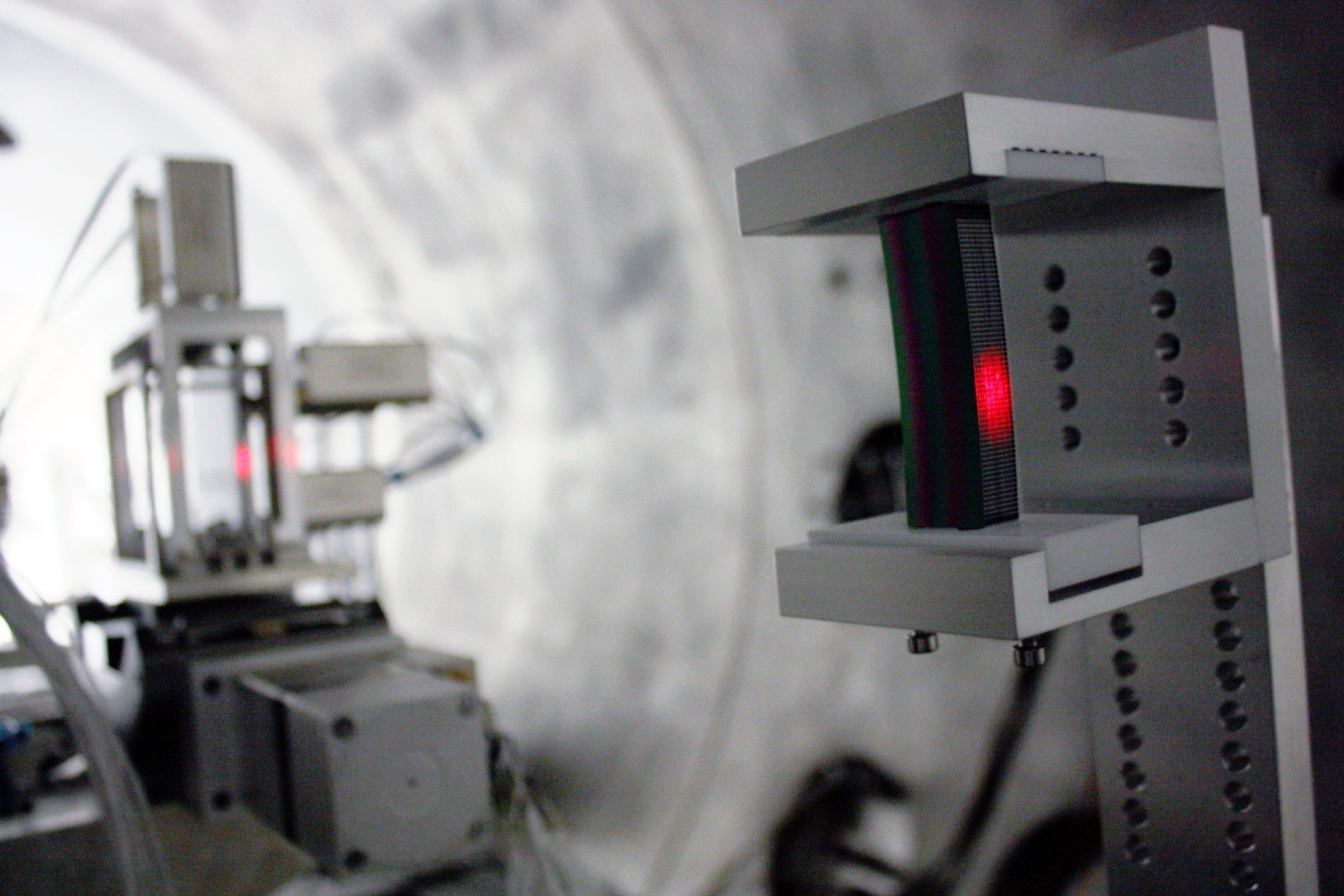}
   \end{center}
   
   \caption{ \label{fig:spo} The SPO as viewed from the source direction during optical alignment. The active alignment grating module is visible in the background. } 
   \end{figure}

The spectrograph assembly tested at the PANTER facility consists of a SPO telescope, off-plane reflection gratings, and a CCD detector at the grating focal plane. A schematic of the SPO, grating, and focal plane positions within the test chamber is shown in Figure \ref{fig:sch}. An image of the components installed in the PANTER vacuum chamber during initial grating alignment with a laser is presented in Figure \ref{fig:spo} where the SPO stack is closest to the camera and the grating module is visible in the background.

\subsection{Silicon Pore Optics}

SPO have been developed for the past 10 years by a consortium led by cosine Research, and have become the main technology for the X-ray mirror of the Athena mission \cite{Willingale13}. SPO make use of industry standard super polished silicon wafers. These wafers are first diced into mirrors plates of the desired rectangular shape, then each plate is wedged using a thin deposition of material on each side so that a focusing optics is formed when multiple plates are stacked onto a conical mandrel. Before being stacked, the plates are ribbed, leaving a thin membrane used to reflect the X-rays and a number of ribs that are used to bond to the next plate. This results in pores in the SPO stack, through which the X-rays can reflect and travel to the focal plane detector. If necessary to meet science requirements, plates can be coated to increase their reflectivity. An SPO stack is very stiff, light-weight, and the stacking process is such that the figure of the mandrel is reproduced with high fidelity so that by combining two stacks in a mirror module, a high resolution imaging system can be built.

For this campaign, a single SPO stack was built. The stack consists of 13 plates, with radii between 450 and 439~mm, width of 66~mm, and length of 22~mm and its geometry approximates that of a parabolic reflector, and t. With a focal length of 8~m, the wedge on each plate was tuned to deliver the required 10~arcsecond ($''$) change in incidence angle between consecutive plates, resulting in a confocal system. For on-axis measurements, the incidence angle is of the order of 1.5$^{\circ}$.

\begin{figure}[t]
   \begin{center}
   \includegraphics[width=.7\textwidth]{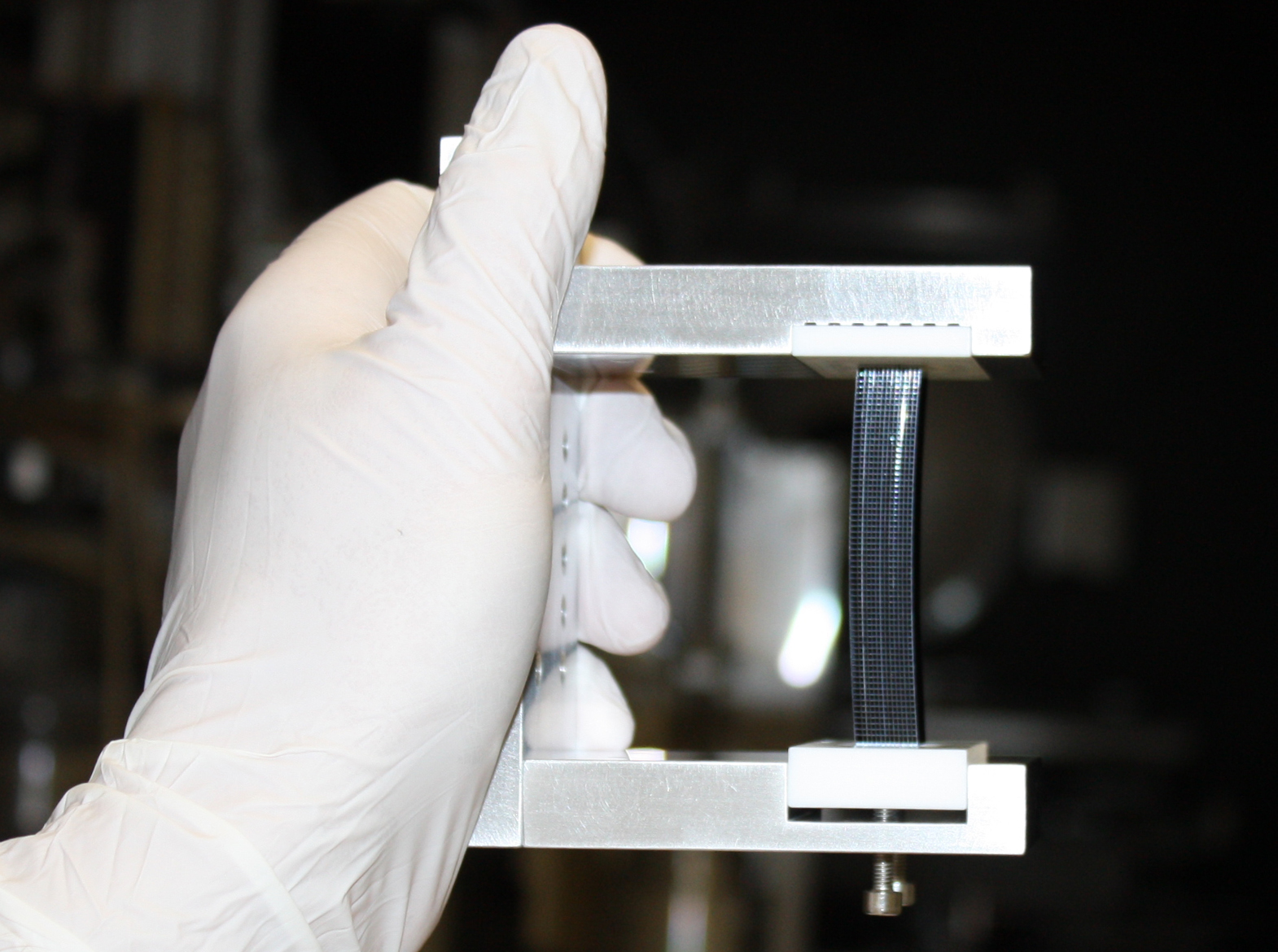}
   \end{center}
   
   \caption{ \label{fig:spo_close} The SPO module prior to installation in the PANTER chamber. } 
   \end{figure}

The SPO stack is shown in Figure \ref{fig:spo_close} prior to installation in the PANTER vacuum chamber. Due to time and budget constraints, the SPO module for this test was shaped using a simple Aluminum mandrel rather than one made of high quality polished fused silica. Therefore, it is important to note that the properties of this SPO module, while qualitatively similar to those in the planned Arcus design, are not characteristic of the state of the art in SPO manufacturing.

\subsection{Gratings}
\begin{figure}[t]
   \begin{center}
   \includegraphics[width=.4\textwidth]{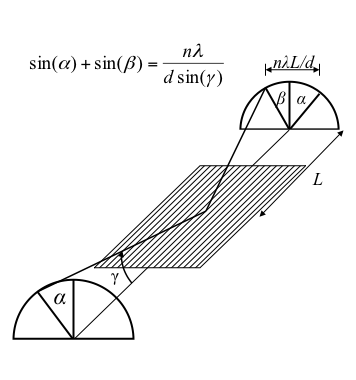} 
   \end{center}
   \caption 
   { \label{fig:grating_geom} Geometry of the off-plane grating mount\cite{McEntaffer13}} 
   \end{figure}

A diagram of the off-plane grating geometry is shown in Figure \ref{fig:grating_geom}. In the off-plane mount, light that is incident onto the gratings at a grazing angle and roughly parallel to the groove direction is diffracted into an arc. The diffraction equation for the off-plane mount is:
\label{equ:grating}
\begin{equation}
\sin \alpha + \sin \beta  = \dfrac{n \lambda}{d \sin \gamma}
\end{equation}
where $\gamma$ is the polar angle of the incident X-rays defined from the groove axis at the point of impact, $d$ is the line spacing of the grooves, $\alpha$ represents the azimuthal angle along a cone with half-angle $\gamma$, and $\beta$ is the azimuthal angle of the diffracted light. The grooves are radially ruled such that the spacing between adjacent grooves decreases towards the focus to match the convergence of the telescope. 

\begin{figure}[t]
   \begin{center}
   \includegraphics[width=.7\textwidth]{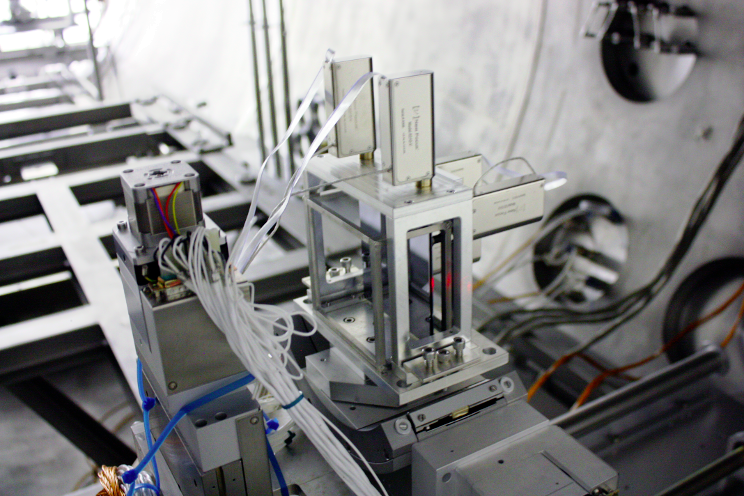} 
   \end{center}
   \caption 
   { \label{fig:aam} The AAM as viewed from the X-ray source direction. The top two picomotors control the active grating yaw while the three motors on the face of the grating (image right) actuate roll and pitch. } 
   \end{figure}

For this test, two gratings were actively aligned together to demonstrate a technique for aligning nested diffraction gratings in an Active Alignment Module (AAM). The AAM consists of slots for the grating wafers and an exterior skeleton into which 5 picomotors are mounted in order to align sequential gratings to an initially installed fixed reference grating. The AAM is shown in Figure \ref{fig:aam}, where two of the five picomotors (used to control grating yaw in this test) are visible on the top of the module. For this test only 1 additional grating was installed and aligned to the reference grating, though the procedure could be repeated to add more gratings as desired. The general installation and alignment procedure is as follows: a reference grating is bonded into the first wafer slot of the AAM. The active wafer is then installed into the adjacent wafer slot with springs between the active wafer and reference wafer surfaces and between the second wafer and the base of the AAM. Finally, the picomotor cage is installed and the 5 picomotors actuate to apply force against the active wafer which is balanced by the surface and base springs. One can now actuate the active grating and finely align it \textit{in situ} using incident X-rays. The step size of each picomotor is approximately 30~nm, which translates to angular step sizes of $\sim$0.1$''$ in roll, pitch, and yaw.  A detailed overview of the active alignment is given by Allured et al. \cite{AlluredIP}.

\subsection{Detectors}
\begin{figure}[t]
   \begin{center}
   \includegraphics[width=.7\textwidth]{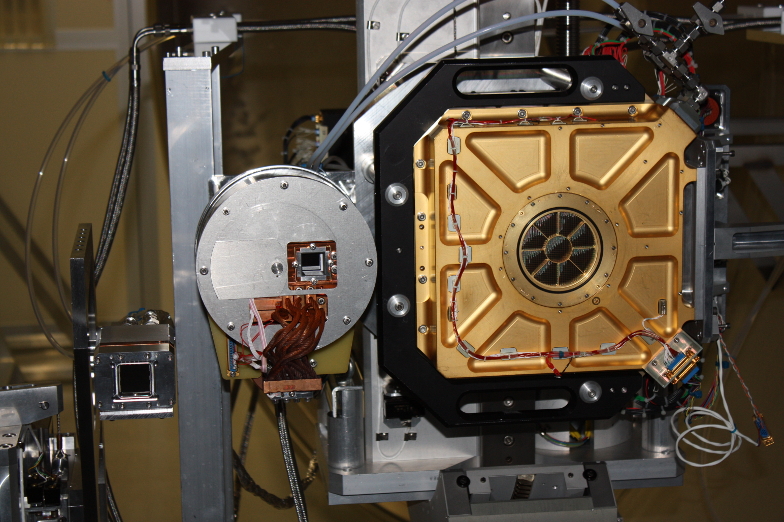} 
   \end{center}
   \caption 
   { \label{fig:det} The X-ray detectors at the PANTER facility. From left to right: PIXI, TRoPIC, and PSPC.  } 
   \end{figure}

Three detectors were in use at the focal plane in the PANTER chamber during these test, TRoPIC, the ROSAT Position Sensitive Proportional Counter (PSPC\cite{Briel86}), and PIXI. TRoPIC is a single photon counting detector with  75~$\upmu$m pixels operated in frame-store mode. TRoPIC is a prototype of the eROSITA detector and is identical apart from its smaller pixel format of 256$\times$256 compared to 384$\times$384 \cite{Meidinger09}. The PANTER facility also has a spare of the ROSAT PSPC detector which we utilized for macro imaging of orders and for rough alignment due to its large active area (80~mm diameter) though relatively course spatial resolution of $\sim$250~$\upmu$m\cite{Pfeffermann87,Pfeffermann03}. PIXI is a Peltier and water cooled Princeton Instruments PI-MTE-1300B integrating in-vacuum CCD with 20~$\upmu$m pixels in a 1340$\times$1300 format \cite{Burwitz13}. All three detectors are shown in Figure \ref{fig:det}. PSPC and TRoPIC are mounted onto the same translation stages and were able to image the 0 and $\pm1^{\rm st}$ orders. PIXI was mounted on a separate vertical translation stage sharing the other movements with TRoPIC and PSPC and was able to reach negative orders. In this paper we focus on measurements taken with the PIXI detector.

\section{PANTER Testing}
\label{sect:exp}

To characterize the spectral resolving power of the grating spectrograph assembly, the gratings and SPO were installed into the detector chamber of the PANTER test facility. The PANTER beamline consists of a multi-target electron impact X-ray source at the head of a 1~m diameter 120~m long vacuum chamber. The beamline ends in a 3.5~m by 12~m test chamber which easily accommodates the SPO, gratings, and detectors.

The gratings were mounted into the PANTER chamber within the AAM with six degrees of freedom relative to the SPO and CCD. The SPO light is incident on the gratings at an angle of 1.5$^{\circ}$. This incidence angle was set using the separation between the SPO focus and the zeroth order reflection of an optical laser mounted at the head of the beamline and has an uncertainty of approximately 1~cm over the 8~m throw ($\sim$4$'$). The chamber was then evacuated, and the PSPC was used find the various diffraction orders and to initially zero the reference grating yaw. Rough alignment was accomplished with PSPC followed by fine alignment with PIXI and TRoPIC. The active grating was aligned to the reference grating by manipulating the AAM picomotors on the active grating (as discussed in \S \ref{sect:telescope}) which allowed for independent roll, pitch, and yaw adjustment.  Once both gratings were aligned, we bonded the active grating in place and remeasured the alignment of the gratings using the PIXI detector.  

\section{Reduction}
\label{sect:reduction}
This section outlines the reduction steps taken to extract LSF measurements from the CCD images taken at PANTER. For the scope of this paper, we focus on images taken with PIXI after the active grating was aligned and bonded into place.

The preprocessing of the images is as follows:
\begin{enumerate}
\item The dark frame is subtracted from the integrated image. The dark frame being a 1340$\times$1300 array where each pixel is the median value of $N$ dark exposures taken consecutively.
\item The background distribution is determined by fitting a normal distribution to the pixels in the non-illuminated edges of the image.
\item The dark-corrected images are thresholded by setting to zero any pixels which fall below 3 times the background $\sigma$ value.
\end{enumerate}
The result of preprocessing is an image whose pixels represent integrated ADC counts.

To fit the LSF of each spectral line, the image is first cropped in $x$ and $y$ around the spectral line. A rotation is then applied to the image via a rotation matrix to account for misalignment with respect to the CCD. The rotation angle is found by performing a least squares fit of the form $y=mx+b$, where $y$ is in the dispersion direction, is applied to the image weighted by each pixel value. This fit is applied for an array of rotation angles, yielding a relationship between the fitted slope and the rotation angle. The best rotation is found as the intercept of a line fit to the rotation angles versus the slope at each angle. This angle is the amount by which the image should be rotated to minimize the slope of the spectral line thereby aligning it with respect to the image x and y axis.

\begin{figure}[t]
   \begin{center}
   \includegraphics[width=.5\textwidth]{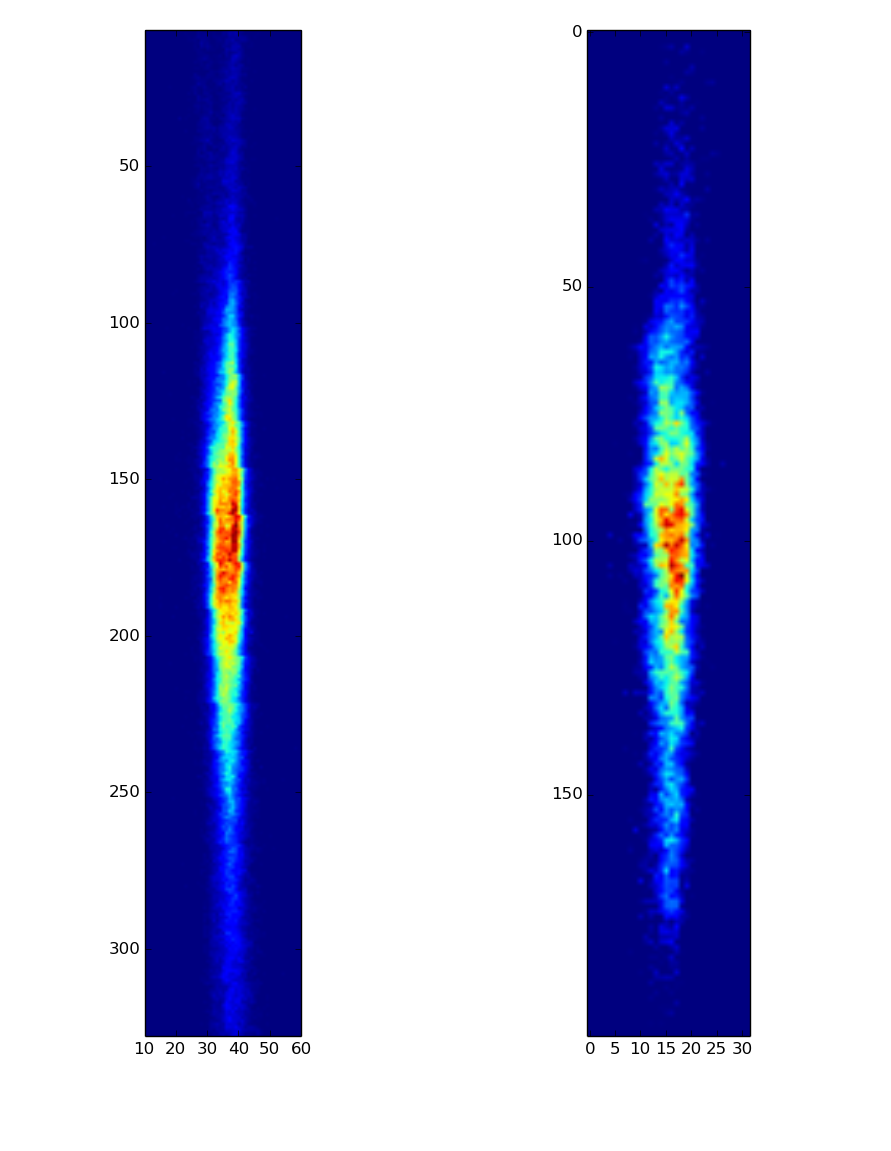} 
   \end{center}
   \caption 
   { \label{fig:CCDims} (Left) PIXI image of the SPO focus; (right) PIXI image of the 1$^{\rm st}$ order Mg-K line. } 
   \end{figure}

Figure \ref{fig:CCDims} shows the cropped and rotated CCD images of the SPO focus and of the first order Mg-K (1.25~keV) line. A `V'-like structure is apparent in the SPO focus (Figure \ref{fig:CCDims} (Left)). Similar structure to the SPO focus also appears, as expected, in the grating focus (Figure \ref{fig:CCDims} (Right)). This structure causes the spectral lines to be poorly fit by simple Gaussian models. Therefore, we describe the line width by the half-energy width (HEW) in the dispersion direction. The CCD image is first flattened in the non-dispersion direction and the HEW of the line is calculated from its cumulative distribution function (CDF) where the HEW contains the central 50\% of the line's integrated ADC counts. The CDF is fit with a spline interpolation, allowing the HEW boundaries fall between pixel values. 

\section{Measured SPO and Telescope Line Spread Function}
\label{sect:results}

\begin{figure}[t]
   \begin{center}
   \includegraphics[width=.7\textwidth]{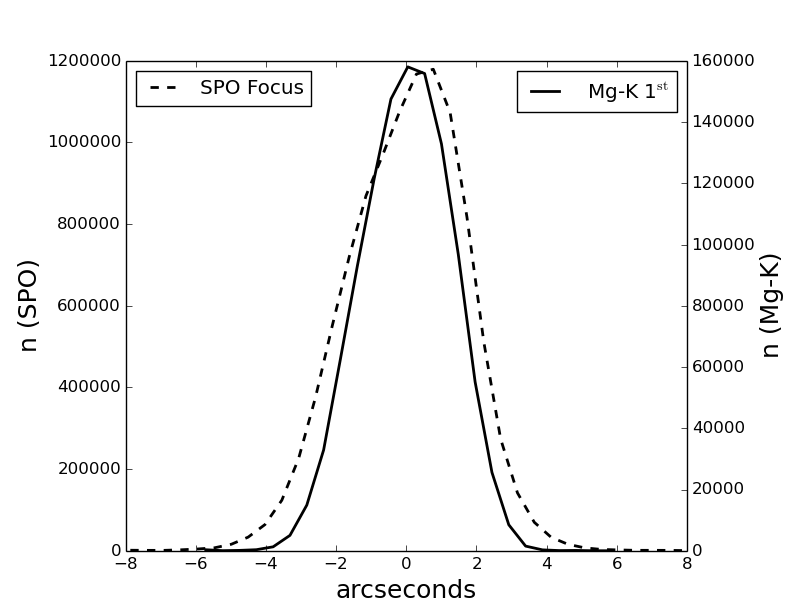} 
   \end{center}
   \caption 
   { \label{fig:spo_lsf} Mg-K 1$^{\rm st}$ order PIXI line profile (solid curve) overplotted with the SPO focus line profile (dashed curve). The HEW are found to be 2.29$''$ and 1.78$''$ for the SPO and Mg-K lines, respectively.}  
   \end{figure}

Figure \ref{fig:spo_lsf} shows the width of the SPO focus in the grating dispersion direction (dashed line) overplotted with the first order Mg-K line (solid line). The Mg-K line is presented due to its natural line width ($\Delta E/E =\mathrm{0.36~eV/1254~eV}$) which is much narrower in dispersion than the width of the SPO focus. The HEW of the SPO focus is found to be 2.29$''$ and the HEW of the Mg-K line is found to be 1.78$''$.

\section{Discussion}
\label{sect:disc}

We have demonstrated a spectrograph composed of silicon pore optics (SPO) and actively aligned off-plane reflection gratings at the PANTER test facility. The half energy width (HEW) of the SPO module is measured to be 2.29$''$ while the HEW of the actively aligned gratings was measured to be 1.78$''$.  We therefore find a narrower line width for the first order Mg-K line than for the SPO focus. A narrower line width for the grating focus could be attributed to the grating sub-aperturing the light from the SPO. While the SPO focus is an integration over all of the SPO plates, the unmasked regions of the gratings only intercept light from approximately two SPO plates. Thus, misalignment between SPO plates will have a larger impact on the width of the total SPO focus than the grating reflection. 

Future development will improve the performance of the telescope. For example, the SPO plates were shaped using a simple Aluminum mandrel. Further iterations will improve on the optic quality by the use of high quality polished fused silica mandrels. We can conclude that the current resolution of the telescope is not limited by the grating resolution or grating alignment, and that higher resolutions can be achieved with further iterations of the mirror assembly.



\bibliographystyle{spiejour}   
\bibliography{refs}
%
%


\vspace{2ex}\noindent{\bf Hannah Marlowe} is a graduate research assistant at the University of Iowa. She received her bachelors degree in astrophysics from Agnes Scott College in 2011. Her current research interests include off-plane X-ray diffraction, X-ray polarimetry, and X-ray spectroscopy.

\vspace{1ex}
\noindent Biographies and photographs of the other authors are not available.

\listoffigures

\end{document}